\documentclass[5p,twocolumn]{elsarticle}
\makeatletter
\def\ps@pprintTitle{%
	\let\@oddhead\@empty
	\let\@evenhead\@empty
	\def\@oddfoot{\rightline{\thepage}}%
	\let\@evenfoot\@oddfoot}
\makeatother

\usepackage{lineno,hyperref}
\usepackage[justification=centering]{caption} 
\usepackage[export]{adjustbox}
\usepackage{subcaption}
\usepackage{amsmath}
\usepackage{amsfonts}
\usepackage[boxed]{algorithm2e}
\usepackage{booktabs}
\usepackage{multirow}
\usepackage{color,soul}  
\usepackage{mathtools, nccmath}
\usepackage{amssymb}
\usepackage{pifont}
\usepackage{hyperref}

\soulregister\cite7      
\soulregister\ref7		 
\modulolinenumbers[5]

\journal{}

\begin{document}

\begin{frontmatter}

\title{Classification of EEG Motor Imagery Using Deep Learning for Brain-Computer Interface Systems}

\author{Alessandro Gallo}
\ead{19305820@student.westernsydney.edu.au}

\author{Manh Duong Phung}
\ead{d.phung@city.westernsydney.edu.au}

\address{Western Sydney University, Sydney City Campus \\
	4/255 Elizabeth St, Sydney NSW 2000, Australia}

\begin{abstract}
	
\textbf{Objective}

A trained T1 class Convolutional Neural Network (CNN) model will be used to examine its ability to
successfully identify motor imagery when fed pre-processed electroencephalography (EEG) data. In theory, and
if the model has been trained accurately, it should be able to identify a class and label it accordingly. The CNN
model will then be restored and used to try and identify the same class of motor imagery data using much smaller
sampled data in an attempt to simulate a live data.

\textbf{Approach}

PyCharm, a Python platform, will be used to house and process the CNN. The raw data used for the training of
the CNN will be sourced from the PhysioBank website. The EEG signal data will then be pre-processed using
Brainstorm software that is a toolbox used in conjunction with MATLAB. The sample data used to validate and
test the trained CNN, will be also be extracted from Brainstorm but in a much smaller size compared to the
training data which is comprised of thousands of images. The sample size would be comparable to a person
wearing a Brain Computer Interface (BCI), offering approximately 20 seconds of motor imagery signal data.

\textbf{Results}

The raw EEG data was successfully extracted and pre-processed. The deep learning model was trained using
the extracted image data along with their corresponding labels. After training, it was able to accurately identify
the T1 class label at 100 percent. The python coding was then modified to restore the trained model and feed it
test sample data in which it was found to recognise 6 out of 10 lines of T1 signal image data. The result suggested
that the initial training of the model required a different, more varying approach, so that it would be able to
detect varying sample signal image data. The outcome of which could mean that the model could be used in
applications for multiple patients wearing the same BCI hardware to control a device or interface.
    
\end{abstract}


\end{frontmatter}


\section{Introduction}
Motor imagery(MI) is known to be the subconscious
link that instigates the interaction between our brain
and our bodies movements. Physical acts are
triggered with intentional and unintentional
thoughts such as pouring a cup of tea (intentional)
or defending against an opponent’s strike, relying on
pure muscle memory and reaction (unintentional or
accidental).

Primarily, the classification of motor imagery
utilises our intentional thoughts with the aim that a
neural network may identify distinct wave form
patterns and class them into their appropriate labels.
These classifications are turned into commands that
can be used to simply apply the accelerator in a
motor car, moving it in a forward direction or
turning the steering at the desired angle. The
outcome of which could mean that a disabled person
may drive a vehicle purely with their minds alone
rather than their body and mind. Companies such as
Daimler the makers of Mercedes Benz vehicles,
have begun research into allowing disabled persons
control their cars interface using only their thoughts.

A Brain computer interface (BCI) is what is used to
interface between the person and the device it is
trying to control. The BCI hardware on the market
today allow for motor imagery extraction from
motor cortex brain signals, that are filtered, and
feature extracted. The unique features in the signal that are born from the motor cortex are what can be
interpreted as a command. The wearer of the BCI
hardware is in turn able to control a device.

Artificial Intelligence (AI) has shown its capability to apply in various problems \cite{malik2018applications, PHUNG2021107376,PHUNG2020106705}.  It also plays the major role in
classifying the extracted signals. Training of a
Convolutional Neural Network (CNN) involves
processing thousands of images of data. This offline
data is fed into the neural network in the form of
training data until it has had enough time to learn at
a realistic rate. An efficient neural network will be
able to differentiate rightly by what is noise and
what is a featured signal of interest.

The outcome of a trained neural network can be
represented as a model. This model can be used to
retrain additional data so that new signals of interest
can be classified, or the model can be used to
classify live or offline data for testing purposes and
ultimately be used to control a device or interface.

Gaps in research into the classification of motor
imagery suggests that there is a reliance to use
offline data for research. There are limited studies
that incorporate online or live data in their papers
and therefore this paper intends to utilise the model
previously implemented, to classify Task1 and with
newly fed samples of EEG signal data to test its
accuracy and usability. Task 1 being the imagery
created when physically opening and closing of both
fists.

\section{Literature Review}

\subsection{Deep learning}
Deep learning falls into three categories; supervised,
unsupervised and semi-supervised learning. When
no labels and classes are known, meaning that the
neural network does not know what its end goal is,
it is a form of unsupervised learning. Machine
learning falls into this category as it relies on
algorithms to work out what it should be looking for
and how. Prior to 2012, the focus of most research was on
unsupervised learning. Semi-supervised learning
requires some data and an algorithm to help it
determine the missing components. Supervised
learning requires input such as labels, classes,
training, testing data and contrastingly, no
algorithms \cite{aa2017deep}.

\subsection{Neural Networks}
CNN’s are considered types of supervised deep
learning architecture models for EEG classification.
Lee et. l \cite{kim2014differences} suggests that they have a
reputation for excellent performance in the field of
image classification that can be used to reduce
‘computation complexities’. Contrastingly though,
Lotte et. al \cite{lotte2018review} argues that their
performances are somewhat held together by their
parameter and architecture combinations. They also
describe the relationship between data size and
architecture, stating that a complex neural
network(NN) with multiple layers require large
datasets for training purposes. They continue to
argue that due to the limited numbers of datasets available for BCI in MI classification, that shallow
neural networks with limited datasets combination
are shown to be more successful.

Other papers suggest that by combining neural
networks, it may produce favourable outcomes, as
each type of model has its distinct advantages over
another {\cite{9406809}. For instance, Sainath et. al \cite{sainath2015convolutional} suggest
that a CNN can help to reduce frequency variations,
Long Short-Term Memory(LSTM) perform better at
temporal modelling and Deep Neural
networks(DNN) are more progressive at mapping
features. Aggarawal and Chugh \cite{aggarwal2019signal} adds that more
recently, CNN’s have been applied for the
classification of multi-classes for motor imagery
tasks by using temporal representations.

The majority of studies point to the lack of datasets
as being the major obstacle in obtaining more
satisfactory results and thereby inhibiting the
research in this field to move forward. Zhang
et. al \cite{zhang2019novel} like many others, have chosen to try and augment their data to try and multiply or artificially
magnify what actually exists. The study used Morlet
Wavelets to transform image signals into three or
four dimensioned tensors, and as a by-product, the
EEG signals are converted to the time frequency
domain.

Both authors in \cite{lotte2018review} and \cite{aggarwal2019signal} agree that more focus should be put on
NN’s to allow them to easily be able to classify
online (non-stationary) data where the sample sizes
would be much smaller and that they should be able
to work with noisier signals.

\section{Methodology}
\subsection{Approach}

A study by Hou et. al \cite{hou2020novel} will be implemented to
validate findings based on their research. They
claim to successfully improve on accuracy scores
attained by other studies in the classification motor
imagery. This research paper will further their
research and use the created and trained CNN deep
learning model to simulate a live testing to
determine if it would be possible to use that model
via a BCI to send a command to a device for the
purpose of mind control.

\subsection{Method of Data Collection}
Acquiring EEG signals is a safe practise according
to (Sanei, Chambers pg. 3 2007), that uses
electrodes of varying types worn using a BCI where
signals are collected in an un invasive manner.
PhysioBank contains multiple databases but the
dataset of interest in this research is from the EEG
Motor Imagery Dataset where the data from 10
patients will be used.

\subsection{Processing the Raw Data}
Utilising an existing EEG database via the
PhysioBank website, raw signal data is to be
imported and processed using Brainstorm software.

Brainstorm is a toolbox that works with MATLAB
as its core to process incoming signals. The raw
EEG data signals are processed to remove unwanted
noise. Such noise can come in the form of high
voltage interference and artefact noise from movement between electrode and the patients head.
Other unwanted signals which is considered noise,
is when the patient is not performing any action or
mentally simulating any action during the signal
recording. Frequencies of interest that correspond
with motor imagery fall between the 5 to 50Hz.This
range of frequencies are what will be kept after
processing has been completed \cite{kim2014differences}.
Processing signal data is an important step that
makes extracting features of interest easier in later
tasks.

\begin{figure}
			\centering
	\includegraphics[width=0.45\textwidth]{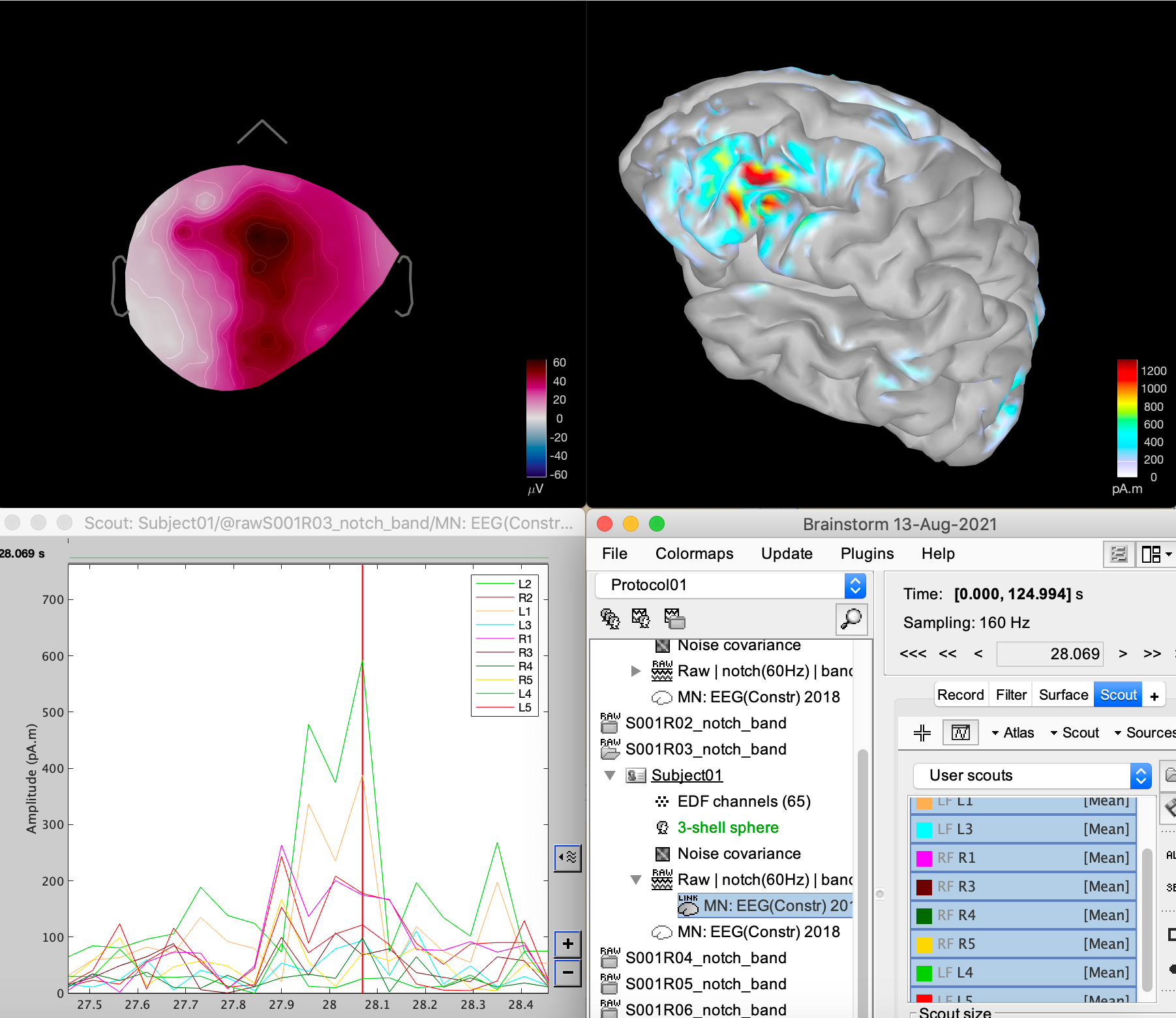}
	\caption{Brainstorm Interface}
	\label{fig1}
	
\end{figure}	

A distinct feature of Brainstorm is that it allows the
mapping of the anatomy of a human brain to the
signals of interest created within the motor cortex
region. In fact, regions of interest are created in the
areas of the brain that exhibit the highest intensity
outputs. This can be seen in figure \ref{fig1} where the
colour is intensified indicating that greater signal
strength is present at scouts L2 and L1 regions.

Morlet Wavelets were used to reconfigure the timebased
signal series into a time frequency-based
system for two purposes, the first being that it is a requirement for a neural network to be able to
distinguish the features of interest using this
rearrangement and second, this method augments
the data to artificially increase the training data size.
The final extraction after pre-processing the signals,
contains timeseries from all scout regions in
MATLAB file format. The files created are then
converted into CSV Excel formatted files for input
into the deep learning model.

\subsection{Deep learning Implementation}
The pre-processed data is organised and split into
training and testing data so that a CNN deep learning
model can learn what patterns of waveforms exist in
the thousands of images presented to it.
Approximately 20,000 images are contained as
training data and 2,000 are for testing. The files are
categorized into the following; Test data, training
data, test labels and training labels. The use of
labelled data indicates that the CNN is a form a
supervised learning, where the model is being
instructed what classes to look out for as it is
learning and again when it is being tested.

\begin{figure}
	\centering
	\includegraphics[width=0.45\textwidth]{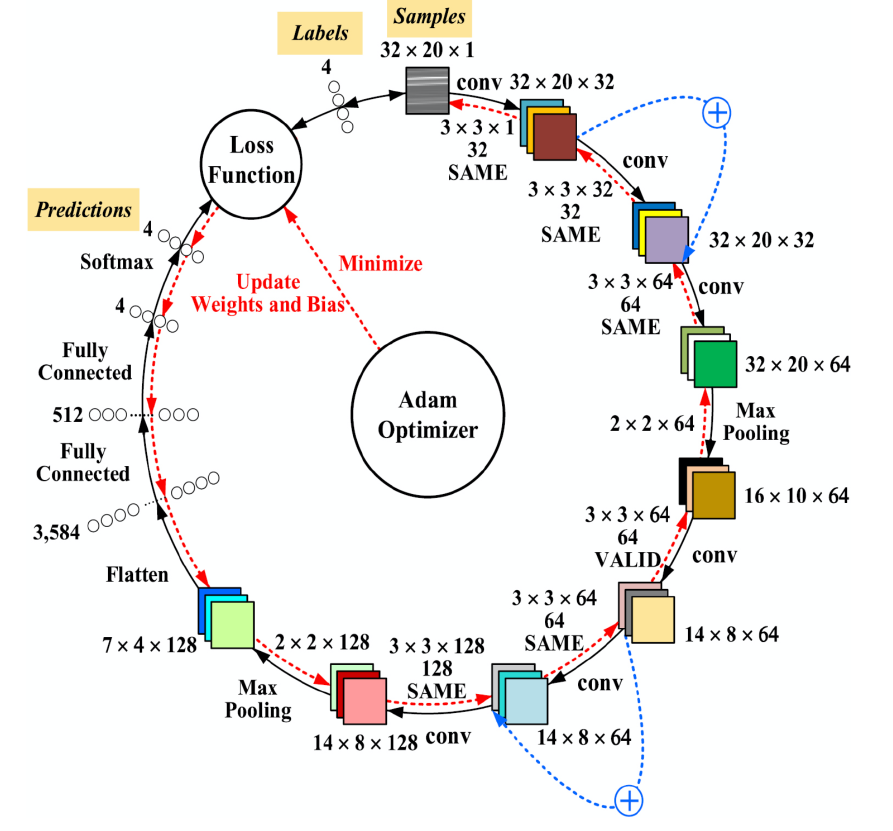}
	\caption{CNN Architecture (Hou et.al 2020)}
	\label{fig2}
	
\end{figure}	

The CNN’s proposed architecture is as implemented
in \cite{hou2020novel}. It consists of 6 convolutional
layers, 2 max pool layers, 2 flatten layers and 1
SoftMax layer. The implemented architecture in
figure \ref{fig2} will be housed in PyCharm and run.
Training of the neural network will be trialled with
differing parameters until a desired result is
achieved. The parameters that will be modified will
be the batch number, the epoch number and possibly
the learning rate. Testing the neural network will
involve modifying the testing data to extract a much
smaller sample and using that data on a previously saved model that has been trained and has produced
adequate results.

\subsection{Method of Analysis}
The proposed method of analysis will be in the form
of quantitative data analysis. Good results produced
by training the CNN model will be indicated by a
high percentage of accuracy. The accuracy in its
ability to be able to distinguish a pattern within the
thousands of images and resolve them to their
rightful class. Graphs will help to visualize the
journey the neural network has undertaken. Such
graphs may indicate a successful convergence
whereby the training accuracy and validation
accuracy will tend to follow each other’s paths. In
contrast the separation of paths would indicate
overfitting, where it could be described when the
CNN has successfully learned from the training data
but fails to transfer its learning when being tested.
Preventing overfitting may require the model to be
made less complicated or by adding dropout layers
which have been shown to effectively prevent this
symptom \cite{srivastava2014dropout}.

\section{Results and Discussion}

\begin{figure}
	\centering
	\includegraphics[width=0.45\textwidth]{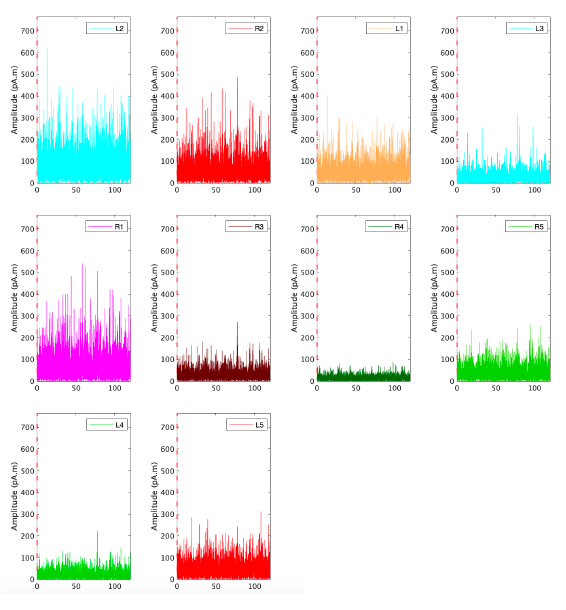}
	\caption{Ten regions of interests}
	\label{fig3}
\end{figure}	

\subsection{Scout Time Series}
All 10 scouts are shown in figure \ref{fig3}. Each scout has
successfully extracted the pre-processed extracted
EEG signals. The amplitudes tend to vary in
intensity between each scout. Scouts R1, R2 and L2
(Figures \ref{fig4}, \ref{fig5}) have experienced more intense spikes
than other scouts.

\begin{figure}
	\centering
	\includegraphics[width=0.45\textwidth]{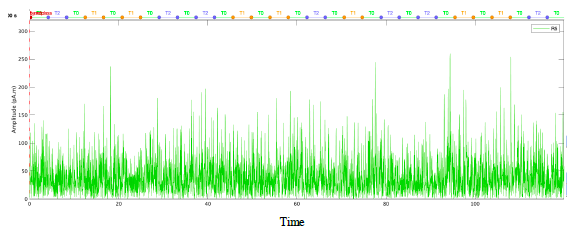}
	\caption{R5 scout signal extraction}
	\label{fig4}
\end{figure}	
\begin{figure}
	\centering
	\includegraphics[width=0.45\textwidth]{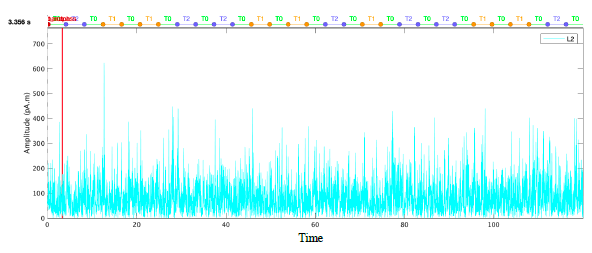}
	\caption{L2 scout signal extraction}
	\label{fig5}
\end{figure}	

The research paper by (Hou et.al 2020) concentrated
their efforts on scout R5 and so the results in this
paper will focus on results based on this scout. In
figure 4 the R5 scout is displayed as a single signal.
Approximately 100 seconds of recording can be
seen along with the task labels on the top of the
figure. Labels T0, T1 and T2 are seen at different
time periods as was recorded of the patient. Task T0
in green is the period of inaction and no imagery takes place. Task 2 describes when the patient
imagines opening and closing their left or right fist.

The label classes contained in the R5 scout are found
in the R2 scout also. Over the same time period in
both scouts it seems that the signals are much denser
in the L2 extraction (Figure 5) than that from the R5
scout. This could possibly affect the neural networks
ability to recognise a pattern from the signals and is
possibly one of the reasons that Hou et. al \cite{hou2020novel}
chose to concentrate their research on the R5 scout
signals of interest.

\subsection{Morlet Wavelets}

\begin{figure}
	\centering
	\includegraphics[width=0.45\textwidth]{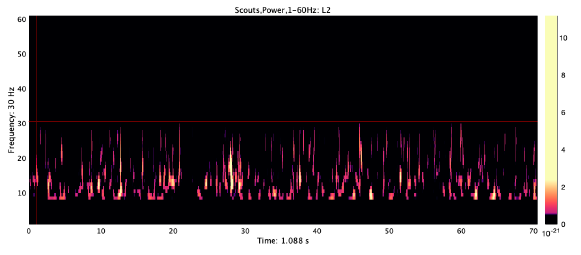}
	\caption{R5 Wavelets}
	\label{fig6}
\end{figure}	

Morlet wavelets were used to extract the time
frequency maps from the scouts as seen in Figure \ref{fig6}.
The frequencies represented, range between 8 and
30 Hz which fits into the range of frequencies that
stem from motor imagery. These features of interest
are then finally extracted into a format recognisable
by the convolutional neural network.
	
\subsection{Training of the CNN} 
Using the PyCharm platform, the Python code was
executed using training data, training labels, testing
data and testing labels. All of which originate from
the R5 scout region that is used as the main source
for training.

\begin{figure}
	\centering
	\includegraphics[width=0.45\textwidth]{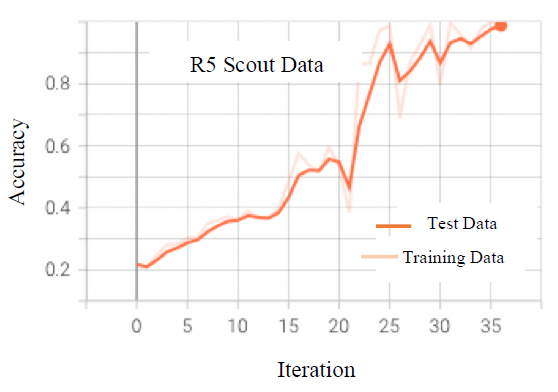}
	\caption{Training Results (Accuracy is
		represented as detection probability)}
	\label{fig7}
\end{figure}	

The CNN was trained with varying parameters until
the training and testing accuracy successfully
converged to produce the results in figure \ref{fig7}. The
results indicated that a class T1 could be recognised
at 100 percent testing accuracy. These results matched
the results obtained in \cite{hou2020novel}.

\subsection{Testing the Trained Model}
This research paper set out to also further the
implementation in \cite{hou2020novel} by then saving
the trained CNN model and using it to test it against
pre-processed sample data. The test data contained
only one image as opposed to approximately 2,000
images when initially training the neural network.
This test would simulate what live data being fed
into an already trained model would look like. This
live data would be equivalent to a person wearing a
BCI device attached to computer waiting for a
command.

\begin{figure}
	\centering
	\includegraphics[width=0.45\textwidth]{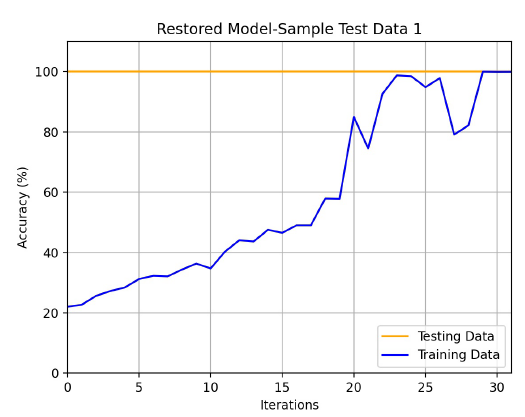}
	\caption{Restored Model}
	\label{fig8}
\end{figure}	

Figure 8 demonstrates the outcome of training the
restored model with sample test data 1.
The accuracy remains at 100 percent for the duration
of the 32 iterations. Meanwhile the training
successfully converged with the testing accuracy.
Although training the restored model was not
necessary, it was reassuring to see the accuracy
remain constant over the duration proving the
restored model’s ability to recognise what it was
trained to do over a given time.

To be able to determine the restored model’s ability
to accept and recognise different samples
individually, a number of tests were performed
using the image data in figure \ref{fig10}. The results shown
in figure \ref{fig9} reveal that when the restored model is
being fed sample data 1-10, it is 60 percent accurate
in identifying the same T1 class. Even though the
trained model had 100 percent accuracy when it was
trained initially using thousands of test data, the
results here would indicate that possibly the training
was not comprehensive enough. This outcome could
be similar to when a student is preparing for an
exam, they would study certain areas of a topic and
then test themselves on the same information scoring highly, but when they actually sit the test,
the questions could have more depth or variance to
them and therefore the student doesn’t score as
highly because they haven’t varied their studies. In
terms of the trained model, it most likely indicates
that more comprehensive training of the neural
network is required in order for it to perform better
against various samples of data.

\begin{figure}
	\centering
	\includegraphics[width=0.45\textwidth]{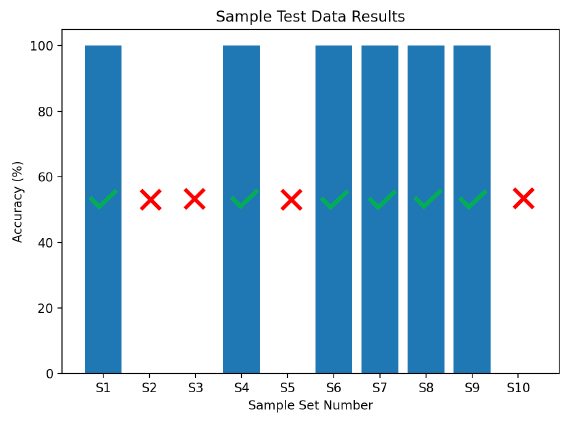}
	\caption{Sample Data Results}
	\label{fig9}
\end{figure}	

\begin{figure}
	\centering
	\includegraphics[width=0.45\textwidth]{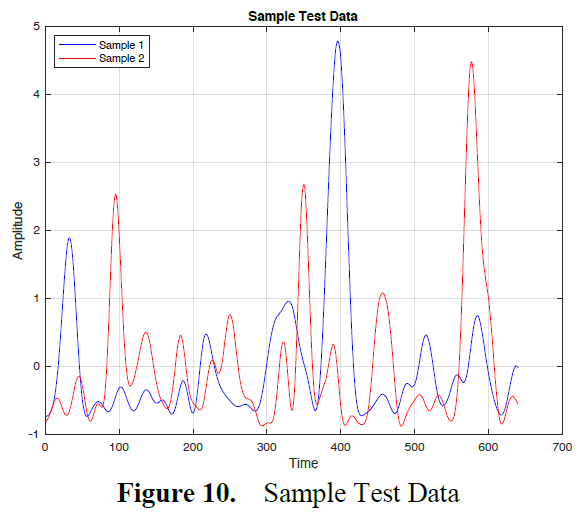}
	\caption{Sample Test Data}
	\label{fig10}
\end{figure}	

The samples plotted in figure \ref{fig10} are samples 1 and
2. The patterns are offset, and amplitudes do differ
but there are similarities in the pattern waveform. The main difference which is quite noticeable is the
added last spike in sample 2. This was probably
unexpected and therefore unrecognizable to the
trained and restored model thereby excluding
sample 2 from its predictions to be of the T1 class.

\section{Conclusion}
This research paper intended to train a neural
network to identify features of interest and class
them according to their appropriate labels. The
implementation followed a paper by (Hou et.al
2020) up to the point where the training was able to
identify a class of MI. This research paper then
attempted to further their research by restoring a
trained model and using it to classify sample image
data that would simulate live data input. This was an
attempt to challenge the evident research gaps in this
field where offline data is the focus of most of the
research in that area.

The methods used in this paper involved preprocessing
the raw EEG signal data in Brainstorm
and successfully extracting the features of interest
that were then converted into the frequency over
time domain. Using the PyCharm platform, the
CNN was trained until it could accurately class a T1
label. The model was then restored and was fed
sample data to test its ability to recognise what could
be potential live data.

The results indicated that the model would need to
be trained using different and varying parameters so
that it would be able to recognise and class various
forms of sample data. Another method may be to try and reduce the complexity of the CNN’s
architecture as was suggested by Lotte et.al 2018 \cite{lotte2018review}.
Either trials could in turn improve the model’s
ability to produce a higher and more consistent rate
of accuracy, ultimately allowing the CNN model to
be used to control a device or combine with other inputs of human \cite{4586369} to carry out more complicated tasks.

\bibliography{reference}

\end{document}